\documentclass{ws-procs9x6}

\def\etal{{\sl et al.}}
\let\ensm=\ensuremath
\newcommand{\gam}{\ensm{\gamma}}

\def\Pp{{\rm p}}
\def\piz               {\ensm{\pi^0}}
\newcommand{\be}{\begin{equation}}
\newcommand{\ee}{\end{equation}}
\newcommand{\rar}{\rightarrow}
\newcommand{\pdup}{p_\uparrow}
\newcommand{\nd}{\noindent}
\newcommand{\ppdup}{$\Pp + \Pp_{\uparrow} \rar \piz + X$}
\newcommand{\pdupp}{$\Pp_{\uparrow} + \Pp \rar \piz + X$}
\newcommand{\pimpr}{$\pi^- + \pdup \rar \pi^0 + X$}
\newcommand{\PR}[1]{{\it Phys.\ Rev.}\ {\bf #1}}
\def\gev              {\ensm{\,{\rm GeV}}}

\def\gevp             {\ensm{\,{\rm GeV}/c}}
\def\mevm             {\ensm{\,{\rm MeV}/c^2}}

\newcommand{\PhL}[1]{{\it Phys.\ Lett.}\ {\bf #1}}
\newcommand{\SNP}[1]{{\it Phys.\ of Atom.\ Nucl.}\ {\bf #1}}
\newcommand{\Yaf}[1]{{\it Yad.\ Fizika}\ {\bf #1}}
\def\xf{x_{\mathrm F}}

\begin{document}

\title{$A_N$ at small negative values of $\xf$ in the reaction 
\ppdup ~at 70 GeV and universal threshold in inclusive pion production.
\footnote{\uppercase{T}his work is supported by \uppercase{R}ussian 
\uppercase{F}oundation for \uppercase{B}asic \uppercase{R}esearch, 
grant 03-02-16919}}

\author{A.M.~Davidenko, A.A.~Derevschikov, V.N.~Grishin,  
V.Yu.~Khodyrev, Yu.A.~Matulenko, Yu.M.~Melnick, A.P.~Meschanin,
V.V.~Mochalov\footnote{\uppercase{C}orresponding author, 
E-mail:mochalov@mx.ihep.su}, L.V.~Nogach, S.B.~Nurushev, P.A.~Semenov,
A.F.~Prudkoglyad, K.E.~Shestermanov, L.F.~Soloviev,
 A.N.~Vasiliev, A.E.~Yakutin}
\address{Institute for High Energy Physics, Protvino, Moscow region, 142281, Russia}

\author{ N.S.~Borisov, A.N.~Fedorov, V.N.~Matafonov\footnote{\uppercase{D}eceased}, A.B.~Neganov, Yu.A.~Plis, Yu.A.~Usov}

\address{JINR,~Dubna, Russia}

\author{A.A.~Lukhanin}
\address{KhPTI, Kharkov, Ukraine}

\maketitle

\abstracts{The talk continues the series of Single Spin Asymmetry (SSA)
$\pi^0$ inclusive measurements carried out at Protvino 70~GeV accelerator.
The asymmetry in the polarized target fragmentation region 
grows up in absolute value with $\xf$ decreasing
and equals to $(-16 \pm 5)\%$ at $-0.4<\xf<-0.25$.  
The result of the current experiment is in agreement with the universal
threshold in pion asymmetries.
}

The experiment was carried out at the PROZA-M 
experimental set-up \cite{setup40} in U-70 Protvino accelerator 
at 1996.
In previous measurements performed in 
Protvino\cite{proza40}\cdash\cite{nurushev} we observed significant asymmetry
in the reaction \pimpr  ~at the central and the polarized target 
fragmentation regions, whereas the asymmetry in the reaction \ppdup ~in the 
central region was compatible with zero. Present talk is devoted to SSA 
measurement at small negative values of $\xf$ in the reaction \ppdup.
  
\begin{figure}[t]
\centering
\includegraphics[width=\textwidth,height=4.5cm]
{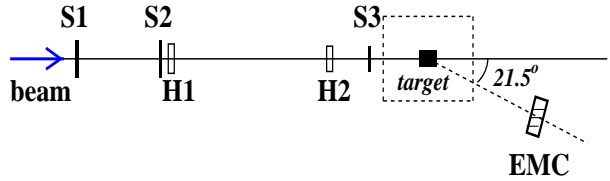}
\vspace*{-1.2cm}
\caption{ Experimental set-up PROZA-M; S1-S3 -- scintillation counters, 
H1-H2 -- two-coordinate hodoscopes, EMC -- electromagnetic calorimeter,
{\it target} - polarized target.
\label{setup}}
\end{figure}

A 70 \gev ~proton beam extracted from the U-70 main ring with
the use of bent crystal had intensity in the range $(3-6) \cdot 10^7$ 
protons/10 sec. cycle. Frozen polarized target ($C_3H_8O_2$) with a
dilution factor $D \approx 10$ operated with 80\% average polarization 
(see Fig.~\ref{setup}).
\gam-quanta were detected by electromagnetic
calorimeter (144 lead glasses) placed 2.8~m downstream target 
at 21.5$^{\circ}$ on respect to the beam direction.
Three scintillation counters S1-S3 and two hodoscopes H1-H2 
were used for beam particle detection and zero level
trigger.  
First level trigger provided the events selection
with the energy deposited in the calorimeter above 1 GeV.
%First level trigger on energy $E$ was used to enrich
%the statistics with high energy events.

The asymmetry was calculated as

\be
A_N=\frac{D(\xf,p_T)}{P_{target}}\cdot A_N^{raw}(\xf,p_T)
=\frac{D(\xf,p_T)}{P_{target}} \cdot 
\frac{n_{\uparrow}(\xf,p_T)-n_{\downarrow}(\xf,p_T)}
{n_{\uparrow}(\xf,p_T)+n_{\downarrow}(\xf,p_T)}
\label{eq:asym}
\ee

\nd 
where {$P_{target}$} -- average target polarization;{$D$} - target 
dilution factor. The azimuthal angle was closed to 180$^{\circ}$, thus
we neglected the SSA dependence on $\cos \phi$.

\begin{center}
\begin{figure}[t]
\begin{tabular}{cp{0.5cm}c}
\includegraphics[width=0.4\textwidth,height=0.38\textwidth]{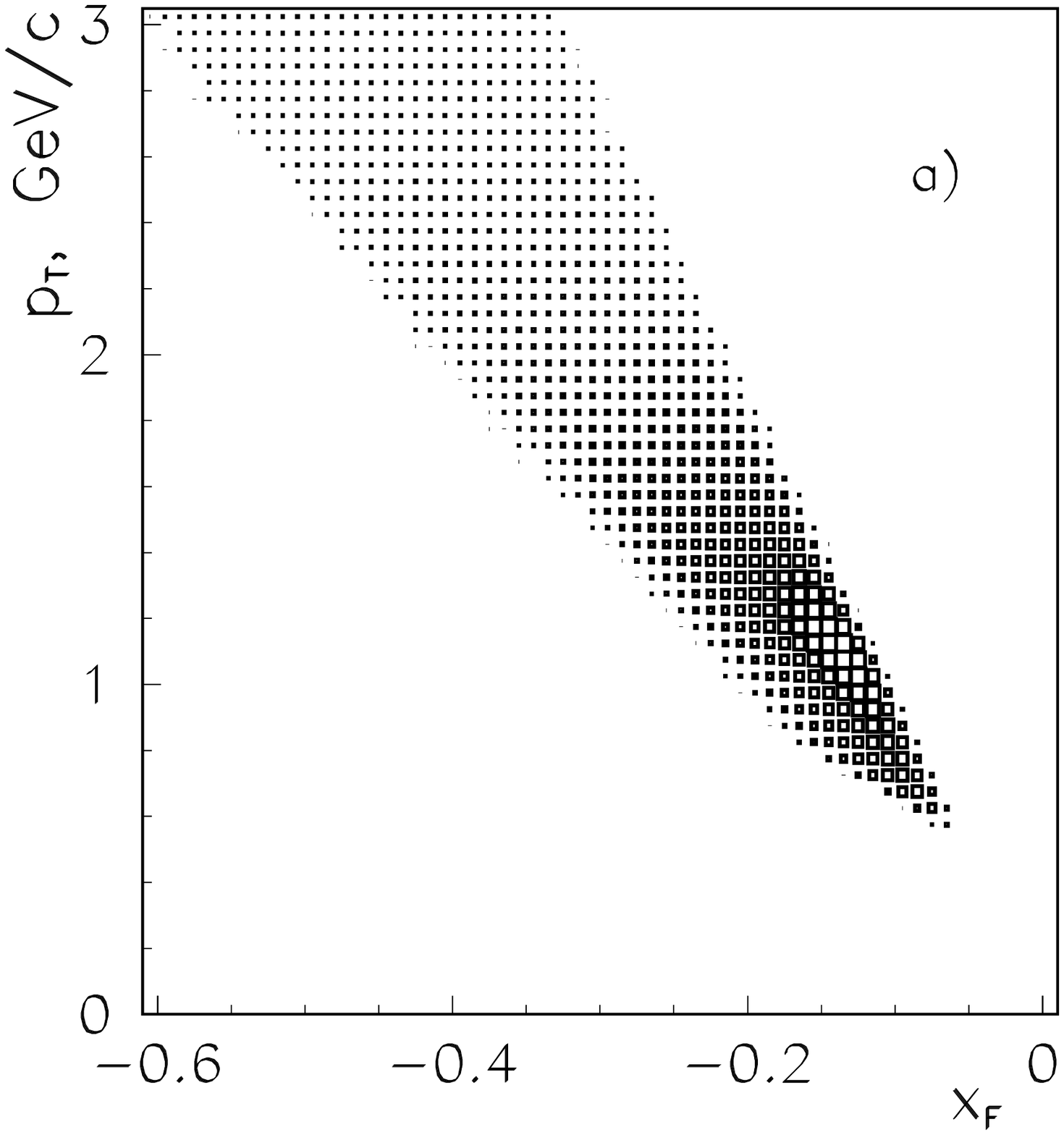} & &
\includegraphics[width=0.4\textwidth,height=0.38\textwidth]{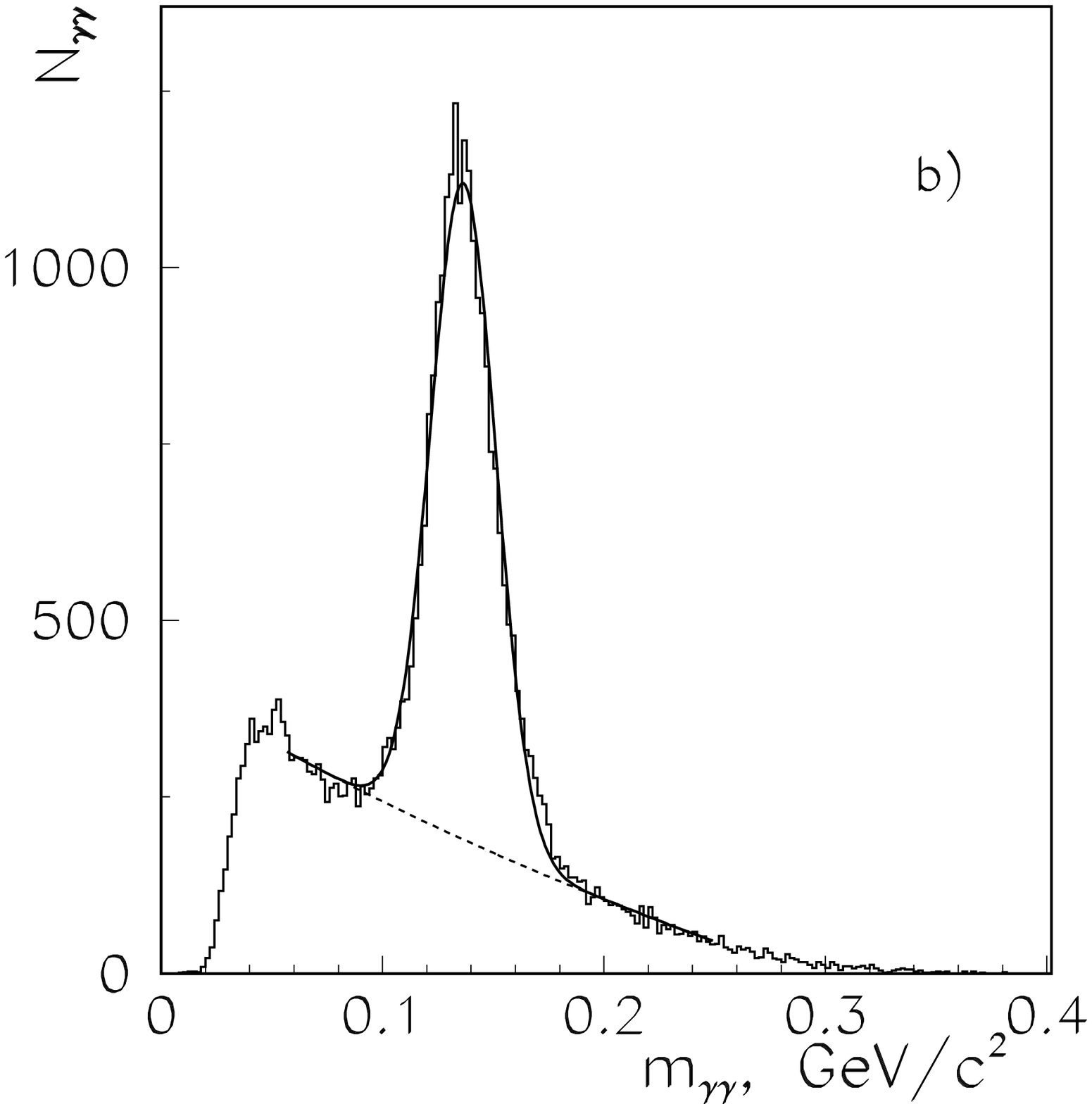} \\
\end{tabular}
\caption{ Two-dimensional distribution on $p_T$ and $\xf$ (a) and Two-gamma
mass spectrum (b)
\label{kinematics}}
\end{figure}
\end{center}

Measurements were performed at narrow solid angle 
(see Fig.\ref{kinematics}), $\pi^0$-mass width was in range 11-17 \mevm ~for
different kinematic regions. 

\begin{figure}[b]
\begin{tabular}{cp{0.5cm}c}
%\vspace*{-0.4cm}
\includegraphics[width=0.45\textwidth,height=3.5cm]
{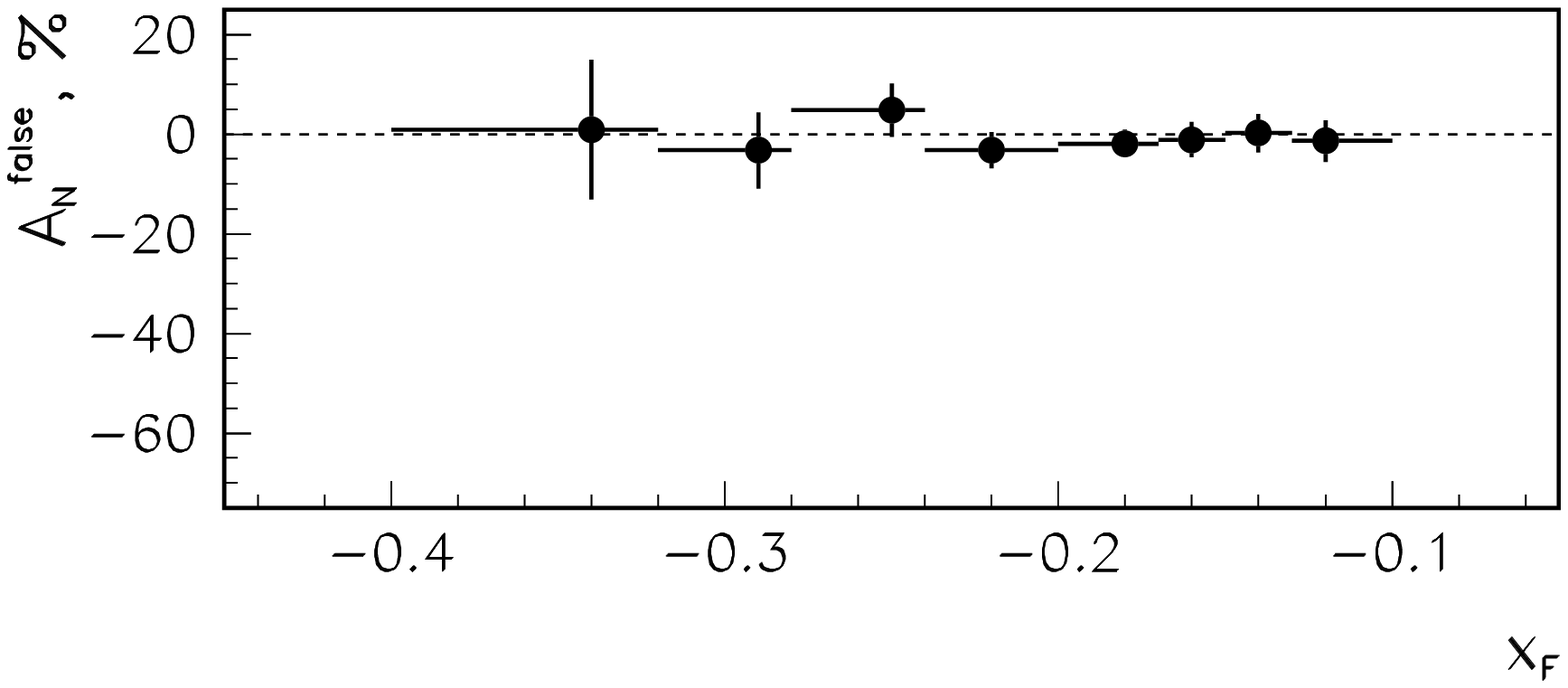}
%\vspace*{-0.7cm}
\includegraphics[width=0.45\textwidth,height=3.5cm]
{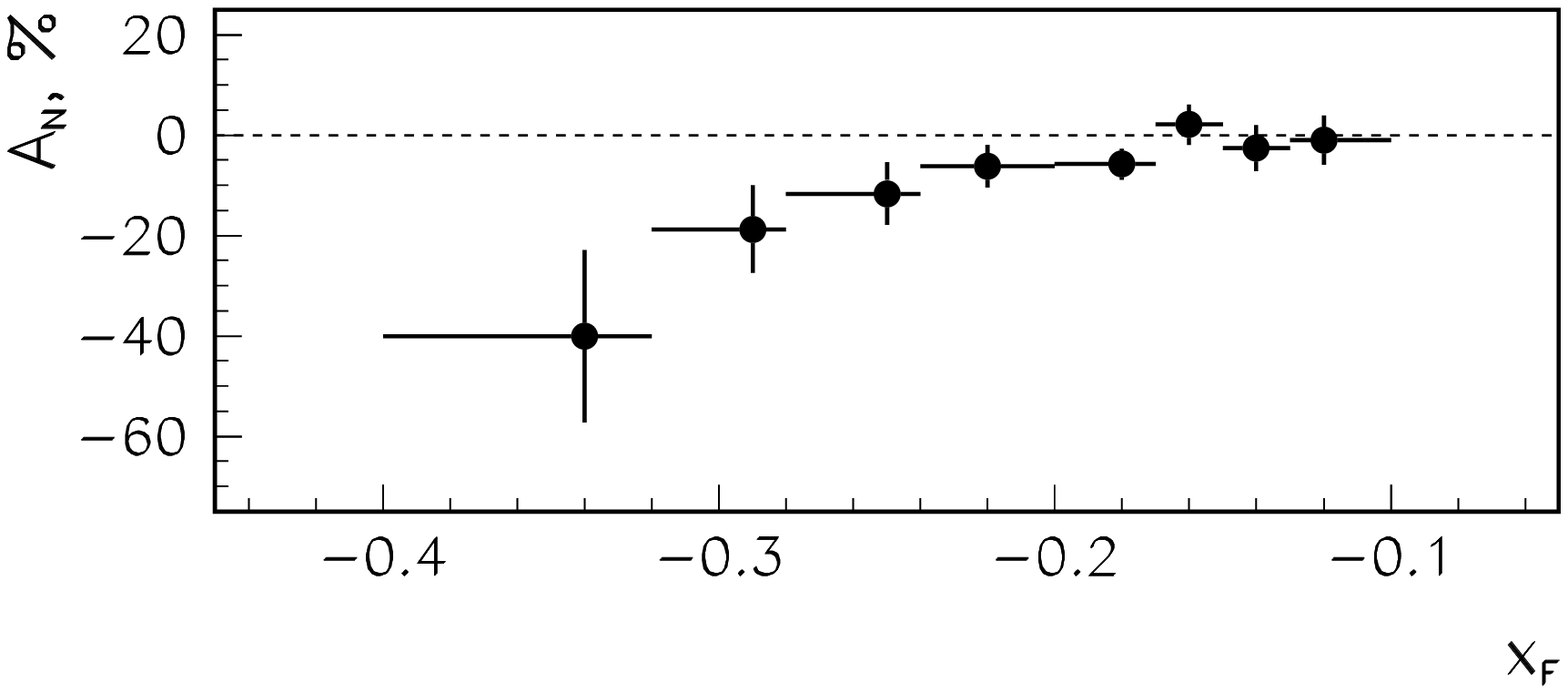}
\end{tabular}
\caption{ False (left) and real (right) $\pi^0$ asymmetry
\label{fig:asym}}
\end{figure}

False asymmetry was compatible with zero (Fig.~\ref{fig:asym}). 
The asymmetry was also close to zero at small absolute values of $\xf$
and grows in magnitude with $|\xf|$ increase. The combined
result $(-16 \pm 5)\%$ is in agreement with other experiments in
polarized particle fragmentation region: 
$A_N=(12.4\pm 1.4)\%$ at E704 (FNAL)\cite{fnal}, $(14\pm 4)\%$ at STAR
(BNL)\cite{star} in the reaction \pdupp and $(13.8\pm 3.8)\%$ in 
the reaction \pimpr ~at PROZA-M (Protvino)\cite{nuclphys40}.

The result is in good agreement with the observation of the universal
threshold $E_0^{cms}$ in SSA if the particle is formed from 
valence quark, which polarization follows the nucleon 
polarization\cite{scal}. The summary of the universal threshold
search for non-zero results is presented in the Table~\ref{table}. All the data
were fitted by the function: 

\be
A_N = \left\{ \begin{array}{ll} 0 & \textrm {, if $E<E_0$} \\
k \cdot (E-E_0) & \textrm {, if $E \geq E_0$} 
\end{array} \right.
\label{threshold}
\ee

The value of $E_0^{cms}$ equals to $(1.5 \pm 0.2)$~\gev 
~for current experiment.

We are planning to carry out new experiment to improve the
accuracy and to measure asymmetry at 
different angles. The detector of 720 cells will be placed on $30^{\circ}$ 
to cover $-0.8<\xf<-0.25$ region.

\begin{table}
\tbl{The search of the universal threshold in the center of mass frame 
for different experiments}
{\footnotesize
\begin{tabular}{|c|c|c|c|c|}
\hline
Reaction & Energy & 
$E^{cms}_0$, GeV& $\chi^2 /N$ 
& $k \cdot (E_{max}^{cms} - E_0^{cms})$, \% \\
\hline
\hline
$\pdup +p \rar \pi^+ + X$ & 13.3 & 
%$65^{\circ}-80^{\circ}$ & 
$ 1.26\pm0.1$ & 0.9 & $52 \pm 6$ \\
$\pdup +p \rar \pi^+ + X$ & 18.5 & 
%$65^{\circ}-80^{\circ}$ & 
$ 1.46\pm0.15$ & 0.85 & $63 \pm 16$ \\
$\pdup +p \rar \pi^+ + X$ & 21.92& 
%$15^{\circ}-25^{\circ}$ & 
$ 1.57 \pm 0.1$ & 0.9 & $68 \pm 6$ \\
$\pdup +p \rar \pi^+ + X$ & 40 &   
% $75^{\circ}-90^{\circ}$ &  
$ 1.64 \pm 0.15$ & &  \\
$\pdup +p \rar \pi^+ + X$ & 200  & 
%$10^{\circ}-20^{\circ}$ &  
$ 1.68 \pm 0.25$ & 1.1 & $52 \pm 5$ \\
\hline
$\pi^- +\pdup \rar \pi^0 + X$ &40 & 
%$80^{\circ}-95^{\circ}$ & 
$ 1.67 \pm 0.15$ & 1.5 & $107 \pm 26$ \\
$\pi^- +\pdup \rar \pi^0 + X$ &40 & 
%$25^{\circ}-30^{\circ}$ & 
$ 1.76 \pm 0.2$ & 0.7 & $36 \pm 14$ \\
$p +\pdup \rar \pi^0 + X$ &24 & 
%$75^{\circ}-90^{\circ}$  & 
$ 1.7 \pm 0.15$ & 0.6 & $334 \pm 165$ \\
$p +\pdup \rar \pi^0 + X$ &70& 
%$35^{\circ}-45^{\circ}$ & 
$ 1.5 \pm 0.2$ & 0.15 & $50 \pm 15$ \\
$\pdup+p \rar \pi^0 + X$ &200 & 
%$10^{\circ}-20^{\circ}$ & 
$ 2.1 \pm 0.3$ & 0.5 & $26 \pm 5$ \\
$\bar{p}_{\uparrow}+p  \rar \pi^0 + X$ &200 &
%$10^{\circ}-20^{\circ}$  & 
$ 0.9 \pm 0.6$ & 0.5 & $13 \pm 4$ \\
\hline
$\pdup +p \rar \pi^- + X$ & 21.92& 
%$15^{\circ}-25^{\circ}$ & 
$ 1.95 \pm 0.1$ & 0.5 & $-87 \pm 11$ \\
$\pdup +p \rar \pi^- + X$ & 200  & 
%$10^{\circ}-20^{\circ}$ & 
$ 2.9 \pm 0.2$ & $<$0.1 & $-51 \pm 6$ \\
\hline
$\bar{p}_{\uparrow}+p \rar \pi^+ + X$ &200 & 
%$10^{\circ}-20^{\circ}$ & 
$ 3.1 \pm 0.5$ & $<$0.1 & $-59 \pm 16$ \\
$\bar{p}_{\uparrow}+p \rar \pi^- + X$ &200 & 
%$10^{\circ}-20^{\circ}$ & 
$ 1.0 \pm 2.2$ & 0.1 & $25 \pm 15$ \\
\hline
\end{tabular}
\label{table}}
%\vspace*{-13pt}
\end{table}

\section*{Conclusion}
--- SSA in the reaction \ppdup was measured at 70 GeV in the kinematic 
region $-0.4<\xf<-0.1$ and $0.9<p_T<2.5$~GeV/$c$.\\
--- Asymmetry equals  zero at  $-0.2<\xf<-0.1$ and 
$0.5<p_T<1.5$~\gevp , and  $A_N=(-16 \pm 5)\%$ at $-0.4<\xf<-0.25$ and $1.3<p_T<3.0$~\gevp.\\
--- The asymmetry starts to grow up at $E_0^{cms}=(1.5 \pm 0.2)$~\gev ~in
agreement with the universal threshold. This behavior can be explained 
inside constituent quark model\cite{const}.\\
--- New measurements will be carried out in next two years
in the \ppdup reaction at the target fragmentation region.

\end{document}